# GIS as a Job Growth Area for IT Professionals

Dr.Timur Mirzoev, Anthony
Moore, Brianna Pryzbysz
Information Technology Department
Georgia Southern University
Statesboro, GA USA

Melissa Taylor
Information Technology Department
Columbus State University
Columbus, GA USA

John Centeno
Information Technology Department
Armstrong State University
Savannah, GA USA

Abstract— As more companies look to capitalize on the benefits of geospatial data, Geographic Information Systems provide an area for growth in the Information Technology job sector in the United States. Careers in GIS require geography, cartography, and IT skills. As the industry grows, candidates with these types of skills that are in demand and are needed to advance the geospatial industry forward. This industry is not generally known as a growth area to many IT professionals, and due to misleading job postings, many candidates may not know their skills are in demand

## I.    INTRODUCTION

In order to understand how Geographic Information Systems (GIS) and geospatial technology can provide job growth for Information Technology (IT) professionals, it is important to understand what GIS and geospatial technologies are and how these technologies work. Geospatial data at its core is similar to any other data with the exception of the spatial element. This spatial element provides a location for the data and x, y coordinates and an elevation z. Just like any other data type, geospatial data requires storage, management, analysis, and dissemination. It is through these requirements that make IT personnel in demand in the GIS industry. As network, database, cloud computing, and development technologies change and grow; the demand for IT personnel to keep the geospatial industry current will grow as well.

Prior to the year 2000 the federal government had been the main user driving the industry. However, since Selective Availability (SA) restrictions on GPS services were lifted in May 2000, geospatial technologies have become more accessible to a wider variety of public industries. Applications such as Google Maps, Facebook Check-in or MapQuest all utilize geospatial technologies. As data collection and extraction increase, the specialized and skilled resource needs will grow. Some of these specialized skills include aerial photography, satellite imagery, understanding and graphical depiction of Census surveys, as well as depicting data collected from businesses such as retail stores. "There is often some level of extraction or processing of relevant information to get data into a desired format or category" [9]. Fig. 1 depicts the shift in GIS users since the 1970s.

The US Department of Labor Employment and Training Administration identified Geospatial Technologies as a high growth industry [23]. Those professionals with a background in professionals. Due to the rapid growth of geospatial technologies into more industries, there are many diverse roles for IT professionals to fill. There are currently highly skilled IT professionals that may fill this void in GIS; however, they may not know their skills are needed. As the field continues to grow, the GIS positions should become more focused and specialized with fewer personnel performing multiple roles. The specialization of job duties will allow more IT personnel without Geography backgrounds to fill positions such as database administrator or data security. This study will elaborate on why GIS is a booming industry, what roles are available, and where to find the jobs.

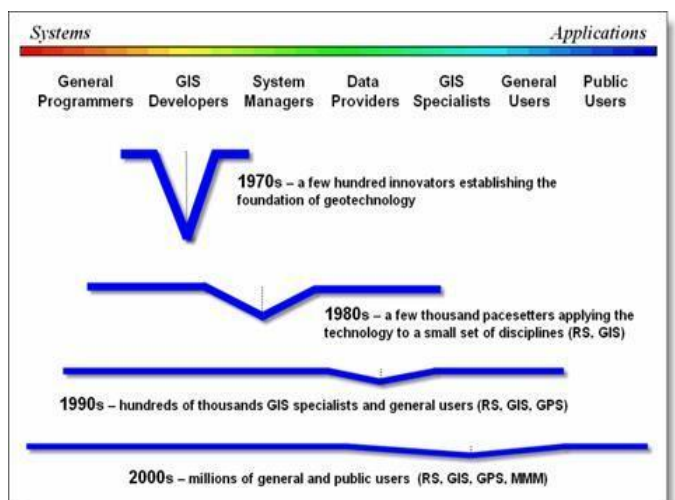

## II.    GEOSPATIAL INDUSTRY OVERVIEW

Much of the growth which is forecasted in the field of GIS is due to governmental and military increase. A recent report,

titled, "GIS Market in US 2012-2016", published by ResearchandMarkets.com in 2013 attributes this to a new found interest in GIS technology over recent years [10]. This rising interest level has led many agencies to increase their annual GIS budget in order to invest into the technology [10]. The United States Armed Forces reports over 400 employees at their Geospatial Center alone [2]. The Geo Center is part of the Army's Core of Engineers Program which has over \$4.5 billion dollar budget for 2015, with a vast amount being spent on GIS [2]. The uses for geospatial technology are so widespread and diverse; the market is growing at an annual rate of almost 35 percent, with the commercial subsection of the market expanding at the rate of 100 percent each year [7].

For IT professionals looking to expand into the GIS industry, it is important to find a proper role. There are many different types of roles to be filled in GIS. Database management is one of the most common IT roles to identify as the GIS database is the heart of the spatial information system. "DBMSs are increasingly important in GIS, since DBMSs are traditionally used to handle large volumes of data and to ensure the logical consistency and integrity of data, which also have become major requirements in GIS" [26]. All of the data that is collected has to be stored somewhere and many commonly used DBMS are used in GIS, such as ORACLE and MySQL. GIS also has a need for developers in the areas of mobile and web applications.

An example of mobile type development is Google Maps, but many other industries are starting to realize the benefits of geospatial data. Typical smart phones come equipped with mapping technology in the form of apps that provide direction to various locations. These apps usually show the route in satellite, hybrid, and 3D modes, and update as a traveler progresses along the route. AT&T provides a family map that allows parents to track their children's whereabouts by using geospatial data received from their cell phones. It also provides a history of their movement patterns that can be used to help determine their possible location in the event they become lost and their device isn't on or working. Along with managing databases and developing applications, server managers are also needed. The U.S. Bureau of Labor Statistics predicts that thousands of new GIS-related jobs will open up by 2020, as more industries recognize the utility of geography-based data management [21].

Even though the geospatial industry has seen tremendous growth, the issue that the geospatial industry has need of professionals with technological backgrounds is not a new issue. Over the past 20, years the geospatial industry has grown so rapidly there has been a shortage in its labor market. In 2006 in an article published by ESRI (Environmental Systems Research Institute), Dr. Duane Marble sought to identify the geospatial workforce, problems facing the workforce, and potential solutions [17]. According to Marble the industry has struggled in identifying itself in the global market and what skill sets comprise its workforce [17]. It is important for the industry to gain understanding of itself and its workforce so that it may better target qualified candidates industry wide.

The geospatial workforce is made up of individuals with diverse skill sets. It is difficult to categorize each position in the geospatial industry with a specific set of skills. Many of the positions held by those in the industry require overlapping skill sets. Identifying skill sets such as those held by information technology professionals and where they fit into the industry will make it easier to target these professionals in the job market.

A study titled "What Skills Does A GIS Analyst Require" assessed the skills needed for each job title along with the salary ranges [25]. In this study, it states "One can often teach a non-spatial person the essentials of spatial information, but it can often be frustrating to have geographers learning programming and databases" [25]. This study also suggests that the salary ranges are quite disparate, ranging from "\$25,000 - \$60,000 per year" [25]. Many technical skills have been defined and are being included in job descriptions such as "Strong Oracle or related RDBMS skills including development skills" [25].

Another study, "Landing a GIS Job and GIS Skills Development in 2013", written by Donoghue and Associates, a consulting firm, discusses the skills employers are specifically looking for specific to the GIS Analyst [25]. These skills include Oracle, MySQL and other database management systems. "Experience with web-application development (.Net, Java, Python, PHP, HTML, JavaScript or Flex)" is also listed as preferred skills [25]. This article also discusses how to obtain certificates or degrees accepted in the GIS field. Donoghue and Associates also recommend starting or participating in a local community of practice.

## III.    METHODOLOGY

Several methods were used when conducting research on GIS positions, beginning with an online search using the internet. As a starting point for the research, focus was put on GIS database, GIS developer, GIS Specialist, and GIS analyst jobs. The goal was to see what types of companies were advertising for those positions and to determine what criteria the job poster used when looking to hire GIS/IT professionals. This would also help to determine how these positions fell across the spectrum in the GIS field. Interviews were conducted with personnel who had experience in the GIS field to get additional insight as to applications that are being used in the business world.

The data was analyzed to determine how companies were advertising for personnel. Specifically, what skill sets were needed, how much education and experience was required, and what the typical duties were for the same job title being advertised. Consistency was compared between companies when they were advertising for GIS staff. For example, to see if a GIS Analyst was defined the same in the majority of the companies looking to fill those positions, or if there were variances in the skills, education and experience, and the actual job duties.

There are many job postings in the GIS field listed on multiple career building websites. In this section, we reviewed the postings specific to GIS on popular job sites as well as GIS-specific job sites and evaluated the job skills and descriptions. Each job posting was assigned a value based on the IT-specific skills required. 1 = Low, 2 = some IT sills and some GIS education preferred, and 3 = high degree of IT skills and GIS specific education required. These criterion are found in Table 1. The effectiveness of each GIS job posting was evaluated using a weighted calculation score which was applied to an overall effectiveness rating for the posting. A comparison chart was assembled to visually compare each of the postings specifically used in this research.

### A. GIS Analyst

A GIS Analyst is responsible for "Reading geographical data, mapping software programming, or even displaying the distance relationships within an entire country". GIS Analysts require skills in math, statistical analysis, spatial skills, as well as creative thinking and communication skills [15].

### B. GIS Data Managers

GIS Data/Database Managers are responsible for designing, configuring, and maintaining the software and databases which store the geographical information. They require database development skills such as SQL as well as analysis skills [19].

TABLE I. JOB POSTING CRITERIA

### C. GIS Developer

A GIS developer is similar to an Information Technology developer in that they both organize and execute activities to design and build applications. A GIS developer, however, does these tasks specific to supporting Geographic Information Systems data. The GIS developers are generally employed by a large corporation or government entity [20].

### D. GIS Specialist

A GIS specialist is a person who uses programs to organize geographic data sets to display information associated to the geography in a digital format easily understood by customers. These specialists manipulate complex geographical data to compare them to other related statistics to create visual maps [24].

Traditionally IT professionals search the internet for jobs using their web browser pointed at the job sites that are the most well-known or the most advertised. The results that are generally returned are jobs posted their by employers or jobs posted on other sites retrieved by web crawlers or search engines. The research for the case in this paper involved searching for GIS & geospatial jobs on the some popular job seeker websites, such as Monster.com, CareerBuilder.com, and Indeed.com as well as some industry specific sites such as GIS Jobs and GIS Clearinghouse.

### IV. FINDINGS

The results returned from one job site versus another vary in quality and quantity. This is due to how the sites obtain the job posts. Monster.com and Careerbuilder.com will generally return fewer job postings for specific search criteria as they rely on jobs to be posted to their site. Indeed.com however is known as a an aggregator site which means job postings on this site have been aggregated or collected from other sites as well as what employers have posted. The differences in job postings totals returned are reflected in Table 2. The GIS specific job sites may return even fewer results as these sites are industry specific and are less well known, these sites require jobs be posted to them as well.

TABLE II. JOB POSTING TOTALS

| Low | Medium | High |
|---|---|---|
| Minimal Information Technology skills<br><br>Limited Education requirements (years not specified)<br><br>Previous Information Technology experience not required<br><br>Previous GIS experience not required | Some Information Technology skills preferred<br><br>Education required, though not specific to GIS (BS or BA)<br><br>Previous Information Technology experience preferred (1-3 years)<br><br>Previous GIS experience preferred (1-3 years) | High degree of Information Technology skills required<br><br>Information Technology education required (BS)<br><br>GIS education required (BS or MS)<br><br>Previous Information Technology experience required (4+ years)<br><br>Previous GIS experience required (4+ years) |

| Job Title Searched | Monster.com | Careerbuilder.com | Indeed.com | | | | |
|---|---|---|---|---|---|---|---|
| GIS Database Administrator | 25 | 228 | 1,618 | GIS Developer | 152 | 86 | 1,046 |
| | | | | GIS Analyst | 1000+ | 138 | 848 |

TABLE III. JOB POSTING COMPARATIVE ANALYSIS

| | Monster.com | | | | | Indeed.com | | | | | CareerBuilder.com | | | | | GJC.org | | | | | GISJobs.com | | | | |
|---|---|---|---|---|---|---|---|---|---|---|---|---|---|---|---|---|---|---|---|---|---|---|---|---|---|
| | A | B | C | D | Total | A | B | C | D | Total | A | B | C | D | Total | A | B | C | D | Total | A | B | C | D | Total |
| GIS Analyst | 2 | 3 | 1 | 3 | 9 | 1 | 2 | 1 | 3 | 7 | 2 | 3 | 2 | 2 | 9 | 3 | 3 | 2 | 2 | 10 | 2 | 3 | 1 | 2 | 8 |
| GIS Data Manager | 3 | 3 | 3 | 2 | 11 | 3 | 2 | 3 | 2 | 10 | 3 | 2 | 3 | 3 | 11 | 2 | 2 | 2 | 3 | 9 | 2 | 2 | 2 | 2 | 8 |
| GIS Developer | 3 | 2 | 2 | 2 | 9 | 3 | 2 | 2 | 2 | 9 | 3 | 2 | 3 | 3 | 11 | 2 | 2 | 2 | 2 | 8 | 3 | 2 | 2 | 2 | 9 |
| GIS Specialist | 2 | 2 | 1 | 2 | 7 | 2 | 1 | 1 | 3 | 7 | 2 | 2 | 1 | 3 | 8 | 1 | 1 | 1 | 2 | 5 | 2 | 2 | 1 | 2 | 7 |

Legend:
A = Technology Skills
B = Education Requirements
C = Information Technology Experience
D - GIS Experience

The findings from each of the job posting searches revealed discrepancies between job requirements for the same position based upon the site the search was performed. These findings are provided in Table 3. Previous Information Technology Experience was the category which scored the lowest on all of the GIS Analyst postings; however, a consistent theme on the GIS Analyst postings showed a requirement of 3 years of GIS experience as well as a Bachelors of Science in the Geography or similar field of study. The lack of previous Information Technology experience or education and focus toward a candidate with a GIS background limits the number of GIS jobs for which skilled IT professionals will qualify.

There are numerous roles an IT specialist can fill in the GIS realm, such as a GIS Analyst. However, depending on how the job is advertised, an IT specialist might not realize his or her skills are a fit for the position. The role of a GIS analyst might mean different things, depending on the company looking for help. For instance, two separate listings on CareerBuilder list completely different requirements for the position. The job requirements for each respectively are:

The ideal candidate would have a "Bachelors or Master's degree in GIS, Geography, Geology and or Environmental sciences (with GIS emphasis)" [5].

The second one lists "B.S. degree in Computer Science, Information system, GIS or related field, and 3+ years' experience working with a GIS system or other enterprise Data Management system" [5].

Without further investigation, job seeking IT professionals might not apply for the first position because they do not see a relevant skill set or educational requirement listed. The GIS Database Manager job postings did show a trend of requiring previous IT experience as well as at least a Bachelors of Science in the Information Technology field. These postings had limited references to the candidate having prior knowledge of GIS systems such as ESRI software. The focus of the postings for the GIS Database Manager revolved around SQL and Oracle knowledge.

The key results of this study presented a significant disconnect between IT professionals searching for GIS roles and GIS employers searching for IT professionals. GIS employers looking to fill a GIS role may focus only on the hiring requirements for GIS professionals, when in reality; an IT professional with exposure to GIS may fulfill the employer's needs in GIS. This seems to be the case for one job reviewed for EKI, an environmental engineer consulting firm looking to fill the role of GIS/Data Analyst. The job posting was made only on the GIS Clearinghouse and list educational requirements of Bachelor's degree in Geography, Environmental Science, Biology, or related field. However, job duties require skills that many IT professionals possess [5]. GIS requires IT more than ever, "GIS users are often enormous consumers of IT infrastructure resources. Consequently IT departments have a large stake and responsibility in GIS implementations" [12]. Spotsylvania, Virginia county government stated that they have a GIS department and an IT department; however, joining the two would result in a better organizational structure and country-wide results [11]. Appendix A contains a synopsis of each posting reviewed during this study.

V.    RECOMMENDATIONS

Based on the research results the IT professionals looking to expand their job opportunities into GIS should include specific keywords in order to find results best suited to their talents. Simply searching for GIS Analyst may not yield as many potential results that the IT job seeker may be qualified to fill. By using specific titles such as developer or database administrator with GIS will produce better results. While all of the job sites returned possible job positions that could be filled by IT professionals, the use of aggregator sites such as Indeed.com returned the most.

The U.S. Department of Labor notes the public is generally not aware of the skills and competencies required to fill the diverse number of GIS positions [23]. It also recommends that the GIS community do a better job of raising awareness of the industry to dispel misperceptions and stereotypes [23]. Another alternative recommendation is for employers to consider alternatives to the traditional pipeline of college education. Some of the recommendations include looking for workforce personnel through apprenticeships and non-traditional labor sources [25].

## VI. Conclusion

The lifting of Selective Availability restrictions in May 2000 has led to an increase in the demand for GIS trained personnel; however, filling those increased demands continues to be a challenge. More and more industries have begun taking advantage of the ongoing business opportunities offered by GIS technology, but they often do so using an incomplete understanding of the education, experience, and overall job requirements of the GIS personnel they are seeking to hire. This general lack of understanding is reflected on job sites such as Monster.com and CareerBuilder.com as there are glaring inconsistencies in how multiple companies advertise for the same job titles and job descriptions. These inconsistencies may mislead prospective candidates into thinking they do not have the requisite skills and/or education necessary for a particular position when they often do. There are always some growing pains with any relatively new technology, and GIS is certainly no exception; however, until public awareness of GIS specific skills and competencies rises, it's likely that there will be continued disconnections within the industry.

APPENDIX A: JOB POSTING REVIEW

| | |
|---|---|
| Posting URL: | https://blm.usajobs.gov/GetJob/ViewDetails/399128600 |
| Posting Title: | Geographic Information Systems Specialist |
| Technology Skills: | N/A |
| Education: | 2 years (36 semester hours) of progressively higher level graduate education leading to a master's degree or master's or equivalent graduate degree in any field that provided the knowledge, skills, and abilities necessary to do the work of the position |
| IT Experience: | N/A |
| GIS Experience | One year of specialized experience preparation, entry & analysis of GIS data; maintain data base; coordinate system user activities |
| Posting URL: | https://careers-soteradefense.icims.com/jobs/5302/database-administrator-%28usda-fsis%29/job |
| Posting Title: | Database Administrator |
| Technology Skills: | Experience with ANSI SQL2 database programming (MS SQL Server, Sybase). Teradata knowledge/experience a plus. Experience developing software applications with the .NET framework (C#, ASP.NET, etc.). Experience with Microsoft Team Foundation Server (TFS). |
| Education: | Bachelor's degree with a major in Computer Science or other major with appropriate programming experience |
| IT Experience: | At least 2 years software design and development experience. |
| GIS Experience | N/A |
| Posting URL: | http://job-openings.monster.com/monster/eaf3e4e9-dc94-479f-9bad-1b4cfd5d15a3?mescoid=1500130001001&jobPosition=6 |
| Posting Title: | Senior GIS Database Administrator |
| Technology Skills: | Highly proficient in administering and deploying open source and commercial off the shelf (COTS) GIS server equipment in support of intelligence organizations. Highly proficient in developing, administering and maintaining Microsoft SQL Server 2012 and/or PostgreSQL in support of intelligence operations |
| Education: | N/A |
| IT Experience: | Five years of technical experience |
| GIS Experience | Ten years GIS database experience within DOD or the Intelligence Community. Expert working knowledge of the discovery, analysis and exploitation of geospatially oriented data and the management of that data within a variety of Open Geospatial Consortium (OGC) compliant applications and databases. Possess high level/expert ability to manage and work with static and dynamic streaming geospatial oriented data |
| Posting URL: | http://job-openings.monster.com/monster/4539358c-9177-47c7-9c35-a305c6713a8c?mescoid=1500130001001&jobPosition=6 |
| Posting Title: | Senior GIS Database Administrator |
| Technology Skills: | Security+ certification required; Network+ and Microsoft Certified Solutions Associate (MCSA) SQL Server 2012 certifications desired. |
| Education: | Bachelor's degree and additional years of general experience in the fields of targeting, intelligence systems architecture and geospatial systems experience are highly desirable. 1-2 years of additional targeted hands on experience can be used in lieu of a Bachelor's degree. |
| IT Experience: | Shall be highly proficient in developing, administering and maintaining Microsoft SQL Server 2012 and/or PostgreSQL |
| GIS Experience | Minimum of ten years GIS database experience within DOD or the Intelligence Community with five years of technical experience and 5 years of leadership experience. Shall be highly proficient in administering and deploying open source and COTS GIS server equipment. Possess an expert working knowledge of the discovery, analysis and exploitation of geospatially oriented data and the management of that data within a variety of Open Geospatial Consortium (OGC) compliant applications and databases. |
| Posting URL: | http://jobs.afacquisitioncareers.com/us/united-states/administrative/jobid7328434-geographic-information-systems-(gis)-specialist |
| Posting Title: | Geographic Information Systems (GIS) Specialist |
| Technology Skills: | N/A |
| Education: | N/A |
| IT Experience: | N/A |
| GIS Experience | Specialized experience includes knowledge of GIS applications and the associated standards, including Spatial Data Standards (SDS), information systems design principles and data modeling concepts to design relational databases. |
| | Skill in developing World Wide Web pages and resolving problems with GIS and inter-related systems<br>Knowledge of AFRC's geobase Program. Knowledge of the principles, practices, methods, and techniques of Geographic Information Systems (GIS).<br>Knowledge of GIS hardware and software systems and their interfaces.<br>Knowledge of concepts, theories, principles, and practices of geospatial sciences.<br>Knowledge of engineering surveying methods, equipment, and techniques.<br>Ability to analyze GIS software and hardware problems, and develop effective and economical solutions.<br>Ability to make mathematical computations using standardized tables and formulas |
| Posting URL: | Http://jobview.local-jobs.monster.com/GIS-Analyst-Job-Grand-Junction-CO-US-147390389.aspx?Ch=denvernewspaper&mescoid=1500140001001&jobposition=8 |
| Posting Title: | GIS Analyst |
| Technology Skills: | N/A |
| Education: | Bachelor's degree in Geography or related field |
| IT Experience: | N/A |
| GIS Experience | 5 years of GIS experience |
| Posting URL: | Http://jobview.monster.com/Geographic-Information-System-GIS-Specialist-Job-Fort-Wayne-IN-US-147940923.aspx?Mescoid=1500131001001 |
| Posting Title: | Geographic Information System (GIS) Specialist |
| Technology Skills: | Knowledge of javascript and Python<br>Knowledge of Silverlight and/or C# languages a plus |
| Education: | Bachelor's degree with two years professional GIS experience OR GIS experience |
| IT Experience: | N/A |
| GIS Experience | Should have at least five years of professional experience in the GIS field<br>Knowledge of GIS Server configuration and maintenance a plus<br>Professional experience with ESRI platform needed<br>Experience with GPS hardware and software a plus |
| Posting URL: | Http://jobview.monster.com/GIS-Analyst-Job-San-Diego-CA-US-147989937.aspx?Mescoid=1500140001001&jobposition=6 |
| Posting Title: | GIS Analyst |
| Technology Skills: | N/A |
| Education: | B.S. in Geography or other area of science or equivalent discipline<br>GIS certificate or demonstrated record of progressing GIS Training/Cartographic Training |
| IT Experience: | N/A |
| GIS Experience | 5+ years experience working with GIS<br>Demonstrated experience developing GIS applications in an enterprise environment using ESRI arcgis<br>Demonstrated experience developing high quality maps that effectively communicate target project data |
| Posting URL: | Http://jobview.monster.com/GIS-Analyst-Job-Tulsa-OK-US-135894395.aspx?Mescoid=1500140001001&jobposition=1 |
| Posting Title: | GIS Analyst |
| Technology Skills: | N/A |
| Education: | Degree in GIS or in a related area (BS/BA preferred) including; Geography, Computer Science, GIS Environmental Discipline or equivalent |
| IT Experience: | N/A |
| GIS Experience | 3 to 8 years experience using GIS applications as an integral part of daily tasks, including use of ESRI products (arcgis 9.x or higher) and Google Earth. |
| Posting URL: | Http://jobview.monster.com/GIS-Specialist-4-Job-Boston-MA-US-147588318.aspx?Mescoid |
| Posting Title: | GIS Specialist 4 |
| Technology Skills: | Familiarity with the tools such as: Microsoft Office Suite, arcgis Desktop, arcgis Server, arcsde, arcgis Online, Esri extensions, Google Earth Pro, Microsoft SQL Server, web and mobile technologies, and basic programming/scripting in HTML5, Python, XML, and Javascript. |
| Education: | B.S./B.A. in GIS, geography, computer science, geology, planning, mathematics, engineering or a related field. |
| IT Experience: | N/A |
| GIS Experience | Four years' experience applying GIS technology with B.S. or 3 years with M.S. |
| Posting URL: | Http://jobview.monster.com/GIS-Specialist-Job-New-York-City-NY-US-147545587.aspx?Mescoid |
| Posting Title: | GIS Specialist |
| Technology Skills: | N/A |
| Education: | 1.A baccalaureate degree from an accredited college including or supplemented by 24 semester credits in computer science or a related computer field and one year of satisfactory full-time computer software experience in computer systems development and analysis, applications programming, database administration, systems programming or data communications; or<br>2.A four-year high school diploma or its educational equivalent and five years of satisfactory full-time computer software experience as described in "1" above; or<br>3.Education and/or experience equivalent to "1" or "2" above. College education may be substituted for up to two years of the required experience in "2" above on the basis that 60 semester credits from an accredited college is equated to one year of experience. In addition, 24 semester credits from an accredited college or graduate school in computer science or a related field, or a certificate of at least 625 hours in computer programming from an accredited technical school (post high school), may be substituted for one year of experience. |
| IT Experience: | See Education |
| GIS Experience | See Education |
| Posting URL: | Http://jobview.monster.com/Oracle-Database-Administrator-Application-Analyst-Job-Westlake-OH-US-148261316.aspx?Mescoid=1500130001001&jobposition=1 |
| Posting Title: | Oracle Database Administrator/Application Analyst |
| Technology Skills: | Experience in PL/SQL, UNIX/Linux scripting, Oracle Forms, RMAN, JAVA, Experience with Oracle Apps version 12, Demantra, & OBIEE,<br>Experience with Oracle databases 8i, 9i, 10g & 11g, Understanding of Tables & table Spaces, Indexes, Views / Materialized views, Triggers, Forms personalization, hardware sizing. |
| Education: | N/A |
| IT Experience: | Experience in PL/SQL, UNIX/Linux scripting<br>Experience with Oracle databases 8i, 9i, 10g & 11g |
| GIS Experience | N/A |
| Posting URL: | Http://wiser.iapplicants.com/viewjob-639213.html?Jb=3&source=6028 |
| Posting Title: | Senior GIS Database Administration / Integration |
| Technology Skills: | Shall be highly proficient in administering and deploying open source and commercial off the shelf (COTS) GIS server equipment in support of intelligence organizations. Shall be highly proficient in developing, administering and maintaining Microsoft SQL Server 2012 and/or postgresql in support of intelligence operations. |
| Education: | Security+ certification. |
| IT Experience: | Five years of technical experience 5 years of leadership experience with the demonstrated ability to effectively manage junior personnel. |
| GIS Experience | Minimum of ten years GIS database experience within DOD or the Intelligence Community.<br>Shall possess an expert working knowledge of the discovery, analysis and exploitation of geospatially oriented data and the management of that data within a variety of Open Geospatial Consortium (OGC) compliant applications and databases.<br>Shall possess high level/expert ability to manage and work with static and dynamic streaming geospatial oriented data. |
| Posting URL: | Http://www.baesystems.jobs/job-geospatial-analyst-8918br?Codes=ID |
| Posting Title: | Geospatial Analyst |
| Technology Skills: | N/A |
| Education: | No Minimum |
| IT Experience: | N/A |
| GIS Experience | 3-10 years of experience required as a Geospatial Analyst. |
| Posting URL: | Http://www.careerbuilder.com/jobseeker/jobs/jobdetails.aspx?Sc_cmp1=js_jrp_jobclick&apath=2.21.0.0.0&job_did=J3G25P76VBDJMJ87T4V&shownewjdp=yes&ipath=QHKV0B |
| Posting Title: | GIS Developer |
| Technology Skills: | Python and javascript knowledge are essential. |
| Education: | 4 Year Degree |
| IT Experience: | 5+ years' experience working with scripting languages. |
| GIS Experience | 5+ years of direct experience working with GIS Tools and applications such as ESRI's arcmap, arccatalog, and arcgis Server. Knowledge of GIS data structures, including topology and also must be familiar with practices for editing GIS data. |
| Posting URL: | Http://www.careerbuilder.com/jobseeker/jobs/jobdetails.aspx?Sc_cmp1=js_jrp_jobclick&apath=2.21.0.0.0&job_did=JHQ78L6NGZZYNZWX8HT&shownewjdp=yes&ipath=QHKV0I |
| Posting Title: | GIS Developer |
| Technology Skills: | Experience with Esri Software Stack (arcgis Desktop, Server, Portal, Online)<br>Experience with Python as a Scripting language.<br>Experience with Python as a OO development language. |
| Education: | Degree from an accredited university in GIS, geography, computer science or related field |
| IT Experience: | 5+ years of related development experience |
| GIS Experience | N/A |
| Posting URL: | Http://www.careerbuilder.com/jobseeker/jobs/jobdetails.aspx?Sc_cmp1=js_jrp_jobclick&apath=2.21.0.0.0&job_did=J3F2HT6929FTRHLDP9N&shownewjdp=yes&ipath=QHKV0B |
| Posting Title: | GIS Developer |
| Technology Skills: | Proficiency with Visual Studio.NET 2010, C# and Python in a service oriented architecture (SOA).<br>Proficiency developing and deploying .NET solutions with IIS.<br>Experience with Software Configuration Management tools. E.g. Team Foundation Server, CVS, Subversion, or Clear Case |
| Education: | N/A |
| IT Experience: | At least 5 year(s) |
| GIS Experience | N/A |
| Posting URL: | Http://www.careerbuilder.com/jobseeker/jobs/jobdetails.aspx?Sc_cmp1=js_jrp_jobclick&apath=2.21.0.0.0&job_did=JHR30M6B0Y32WTBN0X5&shownewjdp=yes&ipath=QHKV0J |
| Posting Title: | Developer - GIS |
| Technology Skills: | Adobe Flex<br>Proficient with Flex Builder IDE<br>Thorough knowledge of Flex framework<br>Deep understanding of Action Script and Adobe Flex builder<br>Exceptional ability to write object oriented code using Flex<br>Demonstrated experience integrating Flex with .Net web services<br>Day to day working with Windows OS, Server 2008+ and IIS Web server |
| Education: | BS in computer science required |
| IT Experience: | Mid-level developer with 3-5 years of work experience |
| GIS Experience | N/A |
| Posting URL: | Http://www.careerbuilder.com/jobseeker/jobs/jobdetails.aspx?Sc_cmp1=js_jrp_jobclick&apath=2.21.0.0.0&job_did=J3F6Y75Z6D2M8194775&shownewjdp=yes&ipath=QHKV0B |
| Posting Title: | GIS Administrator |
| Technology Skills: | N/A |
| Education: | BS in Geographical Information Science, Computer Science or closely related field; |
| IT Experience: | N/A |
| GIS Experience | Four years minimum experience in GIS technologies (ESRI arcgis Desktop, Server and arcsde experience preferred); two years minimum experience in GIS database design, development, implementation and maintenance |
| Posting URL: | Http://www.careerbuilder.com/jobseeker/jobs/jobdetails.aspx?Sc_cmp1=js_jrp_jobclick&apath=2.21.0.0.0&job_did=J3F7VF66KCRB7Y82YDZ&shownewjdp=yes&ipath=QHKV0E |
| Posting Title: | C/Python Developer |
| Technology Skills: | Experience with High-Performance Unix Programming will be an advantage<br>Knowledge of open source queuing systems (rabbitmq) will be an advantage |
| Education: | Bachelor's or Master's Degree in Computer Science, Engineering or related field (or equivalent expertise) |
| IT Experience: | Minimum of 3 years of C development experience or related higher level languages. Along with strong Python and/or Perl development experience |
| GIS Experience | N/A |
| Posting URL: | Http://www.careerbuilder.com/jobseeker/jobs/jobdetails.aspx?Sc_cmp1=js_jrp_jobclick&apath=2.21.0.0.0&job_did=J3H1WZ6CNN67XRCP6RL&shownewjdp=yes&ipath=QHKV0C |
| Posting Title: | Arcsde DBA |
| Technology Skills: | SQL Server, GIS, Arcsde database, Python scripting |
| Education: | N/A |
| IT Experience: | Good understanding of underlying SQL Server database |
| GIS Experience | Extensive hands on experience with deployment of Arcsde in a large enterprise environment (required) |

|  | Advanced knowledge and experience with GIS industry principles, theories and best practices<br>Good knowledge and experience developing physical database design<br>Advanced understanding of GIS data formats, data conversion and data maintenance concepts; Knowledge and experience with spatial data governance |
|---|---|
| Posting URL: | Http://www.careerbuilder.com/jobseeker/jobs/jobdetails.aspx?Sc_cmp1=js_jrp_jobclick&apath=2.21.0.0.0&job_did=J3H6026PZSJYZ6491HW&shownewjdp=yes&ipath=JRKV0B |
| Posting Title: | GIS Solutions Specialist |
| Technology Skills: | SQL |
| Education: | N/A |
| IT Experience: | Enterprise level client server application experience.<br>Network Analyst or Logistics experience.<br>Enterprise Geodatabase design, SQL |
| GIS Experience | 10 years experience working with a GIS system or other enterprise Data Management system; analytical and problem-solving skills<br>Experienced in utilizing arcgis/arcinfo Workstation (arcmap) GIS to QA/QC useable files from client Test program solutions |
| Posting URL: | Http://www.careerbuilder.com/jobseeker/jobs/jobdetails.aspx?Sc_cmp1=js_jrp_jobclick&apath=2.21.0.0.0&job_did=J3J7H65XS51B418G3KM&shownewjdp=yes&ipath=QHKV0I |
| Posting Title: | GIS Analyst |
| Technology Skills: | Applied knowledge and experience with arcgis Desktop, Microsoft Access, SQL, RDMS, Versioned Database Environment, SDE, VB, Geospatial Analysis, Model Builder, Python, Perl, KML, FME, arcims, Microsoft Excel, Information Systems, basic autocad, knowledge of Geographic Positioning Systems |
| Education: | N/A |
| IT Experience: | 6 years of experience required in either PL/SQL, SQL, SQL Plus or SQL Net<br>6 years of experience required in either Perl, ASP, VB, or PHP |
| GIS Experience | 6 years of experience required in either ESRI or arcgis<br>Additional skills not listed above Include: (i) 6 months experience with KML, (ii) 1 year experience with Google Earth and FME or strong combined background with ESRI and scripting can be substituted |
| Posting URL: | Http://www.careerbuilder.com/jobseeker/jobs/jobdetails.aspx?Sc_cmp1=js_jrp_jobclick&apath=2.21.0.0.0&job_did=J3L4KL682ZQ6N0WBRDB&shownewjdp=yes&ipath=QHKV0J |
| Posting Title: | GIS Analyst |
| Technology Skills: | N/A |
| Education: | Bachelor's Degree or higher in geography, or related field of study |
| IT Experience: | N/A |
| GIS Experience | One (1) to three (3) years of professional experience working with ESRI's arcgis (candidates with a Master's Degree and Knowledge of principles, practices and applications of geography, cartography and GIS systems and GPS principles |
| Posting URL: | Http://www.careerbuilder.com/jobseeker/jobs/jobdetails.aspx?Sc_cmp1=js_jrp_jobclick&apath=2.21.0.0.0&job_did=J3L7ZQ6N475HLQFHPNJ&shownewjdp=yes&ipath=QHKV0A |
| Posting Title: | GIS Analyst |
| Technology Skills: | Python |
| Education: | B.S. in Computer Science, Geography or related field and five (5) years of experience working with a GIS system |
| IT Experience: | 3 years of experience with Python programming languages. |
| GIS Experience | Possess at least 2 years of experience utilizing arcgis/arcinfo Workstation (arcmap) GIS to QA/QC useable files.<br>Experience utilizing a variety of software to complete web-programming for GIS applications to include ESRI Desktop Advanced GIS software, Network Analyst extension (ESRI) software, Spatial Analyst extension (ESRI) software, arcserver 10 or 10.1, arcserver Network Analyst extension, arcgis Mobile-software devices communication with GIS mobile devices, ESRI developer network (EDN) developer tools for scripting and software development, Oracle RDMS database and SDE Geodatabase. |
| Posting URL: | Http://www.careerbuilder.com/jobseeker/jobs/jobdetails.aspx?Sc_cmp1=js_jrp_jobclick&apath=2.21.0.0.0&job_did=JHM18474HX19CTQ1B0M&shownewjdp=yes&ipath=JRKV0D |
| Posting Title: | GIS Specialist |
| Technology Skills: | Knowledge of SQL, javascript, or Python a plus |
| Education: | A minimum of a Bachelor's degree in environmental science, geography, GIS, and/or related field. |
| IT Experience: | N/A |
| GIS Experience | Experience with ESRI Products including arcgis versions 10+ and arcgis Online |
| Posting URL: | Http://www.careerbuilder.com/jobseeker/jobs/jobdetails.aspx?Sc_cmp1=js_jrp_jobclick&apath=2.21.0.0.0&job_did=JHS1DL73MHVJ342Y8F4&shownewjdp=yes&ipath=JRKV0F |
| Posting Title: | GIS Specialist |
| Technology Skills: | Microsoft Access |
| Education: | Graduation from a four-year college or university with a major in Geography, Engineering, Surveying, Computer Science or a closely related field; and two years of progressively responsible experience in the design and development of GIS systems and applications or graduation from high school or G.E.D. equivalent and five years of progressively responsible experience in the design and development of GIS systems and applications. |
| IT Experience: | N/A |
| GIS Experience | Experience with Python and arcgis for Server is desirable. ESRI Developer or Enterprise certifications a plus. |
| Posting URL: | Http://www.gisjobs.com/listing.php?Listing=job&id=lbywllbfayaqizbbmbjhrv8m91y088-hxw397tc8rcmzqo12dy |
| Posting Title: | GIS Analyst |
| Technology Skills: | ESRI arcgis 10.x software suite |
| Education: | Bachelor's degree from an accredited college or university which has included major course work in Geographic Information Systems (GIS), Geography, Computer Science, Information Science or Engineering |
| IT Experience: | N/A |
| GIS Experience | Five (5) to Seven (7) years of experience with geographic information systems, including computer graphics and computer hardware digitizing procedures, at least two (2) years which shall have included responsibility for the independent coordination and analysis of computerized geographic survey data |
| Posting URL: | Http://www.gisjobs.com/listing.php?Listing=job&id=s0cmtwhgdd9pttlzjkrfx-5l1vn9l-q3ysdfc_mgdjvk2hk_R2 |
| Posting Title: | GIS Analyst |
| Technology Skills: | arcgis 10.X |
| Education: | Bachelor's Degree in GIS, Geography, Environmental Science or related field |
| IT Experience: | N/A |
| GIS Experience | 2+ years of related experience |
| Posting URL: | Http://www.gisjobs.com/listing.php?Listing=job&id=sgccthxmqmy_5z_257hdiwdoiwwnu4cn_fdsb1sscfsvfrmymt |
| Posting Title: | Geographic Info Systems Specialist |
| Technology Skills: | N/A |
| Education: | Master's Degree in GIS, digital geography, or related discipline. |
| IT Experience: | N/A |
| GIS Experience | Minimum of 3 years experience working in a geospatial environment, preferably in an academic setting. Experience with training for GIS software, including ESRI products. Working knowledge of spatial data formats and related metadata issues. Working knowledge of web mapping applications, such as Google Earth. |
| Posting URL: | Http://www.gisjobs.com/listing.php?Listing=job&id=St4ZH4hFiKuIfSkliBhVrzd7caI3DV-jd29kufd3qxzvek7mad |
| Posting Title: | Geographic Information Specialist |
| Technology Skills: | PHP, Flex/Flash, Python scripts and Modelbuilder is desirable |
| Education: | B.A. /B.S. Degree in Geography, Information Technology or similar discipline required. |
| IT Experience: | N/A |
| GIS Experience | 1 - 2 years providing GIS support to, including internships and advanced course work OFDA or other federal agencies or related organizations |
| Posting URL: | Http://www.gisjobs.com/listing.php?Listing=job&id=vtjla003d2pkvj_F1Kwc44p-5lzh0kvk3ew3tkvwb-hwigq9fm |
| Posting Title: | GIS Analyst |
| Technology Skills: | N/A |
| Education: | Bachelor degree in geographical studies or computer programming |
| IT Experience: | N/A |
| GIS Experience | 3+ years of experience as an Analyst Must have knowledge in cartography/geography and GIS technologies |
| Posting URL: | Http://www.gjc.org/gjc-cgi/showjob.pl?Id=1423940640 |
| Posting Title: | GIS Database Administrator & Developer |
| Technology Skills: | Microsoft SQL, Oracle, postgresql and the upsizing of Microsoft Access databases to enterprise systems |
| Education: | BS/BA in GIS, Geography, Cartography, Computer Science or other related field w/GIS experience. |
| IT Experience: | N/A |
| GIS Experience | Six years of experience in GIS, GIS application development and GIS database administration. At least three years of GIS development or database administration experience supporting a law enforcement mission and at least three years of experience administering databases that hold sensitive data for a national organization or federal agency |
| Posting URL: | Http://www.gjc.org/gjc-cgi/showjob.pl?Id=1426255658 |
| Posting Title: | GIS Analyst |
| Technology Skills: | N/A |
| Education: | Bachelor's Degree in any of the following fields: Geography, Business, Management Information Systems (MIS), Economics, Market Research, Statistics or equivalent analytical degree or experience |
| IT Experience: | N/A |
| GIS Experience | 1 - 3 years in an analytical role specializing in data manipulation, interpretation, report writing, and map generation. 1 - 3 years analytical experience using a Geographic Information System (GIS) and Geospatial analytics |
| Posting URL: | Http://www.gjc.org/gjc-cgi/showjob.pl?Id=1426602109 |
| Posting Title: | GIS Analyst |
| Technology Skills: | Microsoft Office with focus on Excel, Access and PowerPoint |
| Education: | Bachelor's Degree in GIS, Geography, Environmental Science or related field; |
| IT Experience: | N/A |
| GIS Experience | 2+ years of related experience |
| Posting URL: | Http://www.gjc.org/gjc-cgi/showjob.pl?Id=1426622932 |
| Posting Title: | GIS Analyst |
| Technology Skills: | Microsoft Office Professional suite |
| Education: | Bachelor's degree in GIS, Geography, Computer Science, Civil Engineering, or related field |
| IT Experience: | N/A |
| GIS Experience | Minimum two years of technical experience manipulating spatial data |
| Posting URL: | Http://www.gjc.org/gjc-cgi/showjob.pl?Id=1427234856 |
| Posting Title: | GIS Specialist - Enterprise Consultant |
| Technology Skills: | N/A |
| Education: | Bachelor's or master's in geography, GIS, computer science, engineering, or a scientific field |
| IT Experience: | N/A |
| GIS Experience | Minimum of two years of experience with enterprise GIS, arcgis Online and/or Portal for arcgis, arcgis for Server, and system administration |
| Posting URL: | Http://www.gjc.org/gjc-cgi/showjob.pl?Id=1427403321 |
| Posting Title: | GIS Specialist |
| Technology Skills: | Programming experience (Python, Arc Objects, C#, etc) and workflow automation is a plus |
| Education: | N/A |
| IT Experience: | N/A |
| GIS Experience | At least 2-3 Years of practical GIS Analysis and/or IT experience; a Certified GIS Professional (GISP) is a plus |
| Posting URL: | Http://www.gjc.org/gjc-cgi/showjob.pl?Id=1426597006 |
| Posting Title: | GIS Developer |
| Technology Skills: | Silverlight, C#, ASP.NET, Flex, REST, Python, Arcsde, arcserver, arcobjects |
| Education: | N/A |
| IT Experience: | N/A |
| GIS Experience | N/A |
| Posting URL: | Http://www.gjc.org/gjc-cgi/showjob.pl?Id=1426605064 |
| Posting Title: | GIS System Administrator/Developer |
| Technology Skills: | N/A |
| Education: | Bachelor's degree from an accredited college or university with major coursework in Geography, Computer Science, or a related field; and three years of experience in GIS technologies, or a Master's degree and three years of experience in GIS technologies. |
| IT Experience: | N/A |
| GIS Experience | Experience and knowledge of operational support of Enterprise GIS systems and Web application development with ESRI arcgis Server is required. Experience with Google geospatial API is desired. |
| Posting URL: | Http://www.indeed.com/cmp/Finezi-Inc/jobs/GIS-Specialist-21d4de920810a128 |
| Posting Title: | GIS Specialist |
| Technology Skills: | SQL, Python |
| Education: | N/A |
| IT Experience: | Knowledge of Structured Query Language (SQL) scripting language to retrieve data from relational database management systems in MS SQL Server |
| GIS Experience | 5-7 years of GIS Analyst experience utilizing the principles and practices of GIS is required<br>Considerable knowledge and experience with the core ESRI GIS software products, specifically arcgis Desktop, Arc/Info, arcsde, and relational databases within MS SQL Server. |
| Posting URL: | Http://www.indeed.com/cmp/Mississippi-Department-of-Marine-Resources/jobs/GIS-Database-Administrator-f661af40412f7 2ef?Sjdu=qwrrxkrqz3cmx5w-o9jevdlszaj7g0xm71kycvepbkspozuc-bjnq7cgzw1ywanaze-nyiyvkhlv8ha2xrqfoh5fwyxzrduy2t_ pkafzle0 |
| Posting Title: | GIS and Database Administrator |
| Technology Skills: | N/A |
| Education: | A Master's Degree from an accredited four-year college or university |
| IT Experience: | N/A |
| GIS Experience | Seven (7) years experience, five (5) years of which must have included line or functional administrative or advanced technical supervision |
| Posting URL: | Http://jobview.monster.com/GIS-Application-Developer-Job-Saint-Louis-MO-US-148182948.aspx?Mescoid=1500140001001 &jobposition=4 |
| Posting Title: | GIS Application Developer |
| Technology Skills: | Familiarity with programming languages (e.g. ColdFusion, .NET, Python, PHP, Java). |
| Education: | Undergraduate or graduate degree in Computer Science or related field or at least five years of programming experience without a specified degree. |
| IT Experience: | N/A |
| GIS Experience | N/A |
| Posting URL: | Http://jobview.monster.com/GIS-Developer-Geographic-Information-Systems-Job-Kansas-City-MO-US-147563343.aspx?Me scoid=1500140001001&jobposition=5 |
| Posting Title: | GIS Developer (Geographic Information Systems) |
| Technology Skills: | JavaScript, and .Net application development |
| Education: | N/A |
| IT Experience: | Two plus (2+) years' experience developing Web applications |
| GIS Experience | Two plus (2+) years' experience developing GIS applications |
| Posting URL: | Http://jobview.monster.com/GIS-Developer-Job-Washington-D.C.-PA-US-148149923.aspx?Mescoid=1500140001001&jobpo sition=3 |
| Posting Title: | GIS Developer |
| Technology Skills: | HTML and JavaScript |
| Education: | Bachelor's degree or higher in either Computer Sci. (or other related degree) or Geography |
| IT Experience: | N/A |
| GIS Experience | 6+ years of GIS database and application experience |
| Posting URL: | Http://www.indeed.com/cmp/agileassets/jobs/Senior-GIS-Developer-56dbf93af6a64fa1 |
| Posting Title: | Senior GIS Developer |
| Technology Skills: | Oracle |
| Education: | Bachelor Degree in Computer Science or related field |
| IT Experience: | 3+ years' experience and a demonstrated proficiency in working with RDBS (Oracle preferred)<br>3+ years' experience in Web Application development in Java and javascript; OOP is a must |
| GIS Experience | 3+ Years of experience and strong understanding of GIS technology |
| Posting URL: | Https://johnsonmirmiranthompson-openhire.silkroad.com/epostings/index.cfm?Fuseaction=app.dspjob&jobid=995&company_id=16357&version=1&jobboardid=1112 |
| Posting Title: | GIS Applications Developer |
| Technology Skills: | arcgis and Oracle products, data structures, architectures, development tool and environments, and relational database management systems |
| Education: | Master's degree in a relevant discipline such as computer science or geography, or in any discipline but with specialization in GIS technology, may substitute for two (2) years of development experience. Bachelor's degree in a relevant discipline such as computer science or geography, or in any discipline but with a specialization in GIS technology, may substitute for one (1) year of development experience. |
| IT Experience: | Experience in a managed data environment, including relational databases and storage area networks |
| GIS Experience | Six (6) or more years experience with development using functional and nonfunctional requirements within a specified architecture, or with development of server jobs and tools for the maintenance of large, complex enterprise GIS, GIS databases or GIS applications, as appropriate for the specific task assignment; two (2) year's task management and self-supervisory experience including progressive experience in managing complex, difficult tasks |
| Posting URL: | Http://www.indeed.com/viewjob?Jk=6499115a45cf71c9&q=gis+developer&tk=19hjjcqp70n3h1a8 |
| Posting Title: | GIS Developer |
| Technology Skills: | Esri Software Stack (arcgis Desktop, Server, Portal, Online). |
| Education: | Degree from an accredited university in GIS, geography, computer science or related field. Advanced degree preferred. |
| IT Experience: | 5+ years of related development experience |
| GIS Experience | N/A |
| Posting URL: | Http://www.gisjobs.com/listing.php?Listing=job&id=ewqsdn-0Yyn0XYBksbo70J6TvAL6JelOBCuEest2RGRCtgNMh1 |
| Posting Title: | Software Developer / Programmer |
| Technology Skills: | C#, javascript, Microsoft .NET for desktop and web applications, HTML5 and REST web services, RDBMS (Oracle, SQL Server, etc.), Cross-browser compatibility issues |
| Education: | Bachelor degree in Computer Science, with concentration in Programming, Software Engineering or IT |
| IT Experience: | 3-5 years of proven programming or software development experience |
| GIS Experience | 3+ years of hands-on experience and excellent working knowledge of GIS development experience using Esri arcgis Engine and arcgis for Server apis |
| Posting URL: | Https://www1.apply2jobs.com/EOIR/profext/index.cfm?Fuseaction=mexternal.showjob&RID=1405¤tpage=1 |
| Posting Title: | Geospatial Analyst |
| Technology Skills: | N/A |
| Education: | Bachelor's Degree and Master's Degree |
| IT Experience: | N/A |
| GIS Experience | Experience applying geospatial analytical methods, substantive intelligence expertise and logical judgments to solve problems |